\documentclass[prb,showpacs,twocolumn,amsmath,amssymb,floatfix]{revtex4}

\usepackage{graphicx}

\begin{document}

\newcommand{\IJMPB}{{Int. J. Mod. Phys. B} }
\newcommand{\PhC}{{Physica C} }
\newcommand{\PhB}{{Physica B} }
\newcommand{\JS}{{J. Supercond.} }
\newcommand{\IEEEmw}{{IEEE Trans. Microwave Theory Tech.} }
\newcommand{\IEEEas}{{IEEE Trans. Appl. Supercond.} }
\newcommand{\IEEEim}{{IEEE Trans. Instr. Meas.} }
\newcommand{\PR}{{Phys. Rev.} }
\newcommand{\PRB}{{Phys. Rev. B} }
\newcommand{\PRL}{{Phys. Rev. Lett.} }
\newcommand{\IJIMW}{{Int. J. Infrared Millim. Waves} }
\newcommand{\APL}{{Appl. Phys. Lett.} }
\newcommand{\JAP}{{J. Appl. Phys.} }
\newcommand{\JPCM}{{J. Phys.: Condens. Matter} }
\newcommand{\JPCS}{{J. Phys. Chem. Solids} }
\newcommand{\AdP}{{Adv. Phys.} }
\newcommand{\Nat}{{Nature} }
\newcommand{\CM}{{cond-mat/} }
\newcommand{\JpnJAP}{{Jpn. J. Appl. Phys.} }
\newcommand{\PhT}{{Phys. Today} }
\newcommand{\ZETF}{{Zh. Eksperim. i. Teor. Fiz.} }
\newcommand{\JETP}{{Soviet Phys.--JETP} }
\newcommand{\EL}{{Europhys. Lett.} }
\newcommand{\Sci}{{Science} }
\newcommand{\EJPB}{{Eur. J. Phys. B} }
\newcommand{\IJMB}{{Int. J. of Mod. Phys. B} }
\newcommand{\RPP}{{Rep. Prog. Phys.} }
\newcommand{\SUST}{{Supercond. Sci. Technol.} }

\title{Reduction of the field-dependent microwave surface resistance in YBa$_{2}$Cu$_{3}$O$_{7-\delta}$ with sub-micrometric BaZrO$_3$ inclusions as a function of BaZrO$_3$ concentration.}

\author{N. Pompeo}
\affiliation{Dipartimento di Fisica ``E. Amaldi'' and Unit\`a CNISM, Universit\`a Roma Tre, Via della Vasca Navale 84, I-00146 Roma, Italy}
\author{R. Rogai}
\affiliation{Dipartimento di Fisica ``E. Amaldi'' and Unit\`a CNISM, Universit\`a Roma Tre, Via della Vasca Navale 84, I-00146 Roma, Italy}
\author{A. Augieri}
\affiliation{ENEA-Frascati, Via Enrico Fermi 45, 00044 Frascati, Roma, Italy}
\author{V. Galluzzi}
\affiliation{ENEA-Frascati, Via Enrico Fermi 45, 00044 Frascati, Roma, Italy}
\author{G. Celentano}
\affiliation{ENEA-Frascati, Via Enrico Fermi 45, 00044 Frascati, Roma, Italy}
\author{E. Silva\footnote{corresponding author. e-mail: silva@fis.uniroma3.it}}
\affiliation{Dipartimento di Fisica ``E. Amaldi'' and Unit\`a CNISM, Universit\`a Roma Tre, Via della Vasca Navale 84, I-00146 Roma, Italy}

\begin{abstract}
\noindent In order to study the vortex pinning determined by artificially introduced pinning centers in the small-vortex displacement regime, we measured the microwave surface impedance at 47.7 GHz in the mixed state of YBa$_{2}$Cu$_{3}$O$_{7-\delta}$ thin films, where sub-micrometric BaZrO$_3$ particles have been incorporated. As a function of the BaZrO$_3$ content, we observe that the absolute losses slightly decrease up to a BaZrO$_3$ content of 5\%, and then increase. We found that the magnetic-field-induced losses behave differently, in that they are not monotonic with increasing BaZrO$_3$ concentration: at small concentration (2.5\%) the field-induced losses increase, but large reduction of the losses themselves, by factors up to 3, is observed upon further increasing the BaZrO$_3$ concentration in the target up to 7\%. 
Using measurements of both surface resistance and surface reactance we estimate vortex pinning-related parameters. We find that BaZrO$_3$ inclusions introduce deep and steep pinning wells. In particular, the minimum height of the energy barrier for single vortices is raised. At larger BaZrO$_3$ content (5\% and 7\%) the phenomenon is at its maximum, but it is unclear whether it shows a saturation or not, thus leaving room for further improvements.
\end{abstract}


\maketitle

\section{Introduction}
\label{intro}
Applications of type-II superconductors require strong pinning of flux lines to avoid vortex-motion losses. In layered High-$T_c$ Superconductors (HTCS) this is a long-standing issue: the layered structure yields flexible flux lines, the high operating temperature gives rise to large thermally induced vortex creep or even thermally activated flux flow. Up to recent times, it was believed that only columnar defects\cite{civalePRL91} could drastically improve the performances of HTCS in magnetic fields. Thus, the interest was relegated to the most basic studies, due to the impractical method of introducing such defects (heavy ion irradiation). However, in the last years it has been shown\cite{macmanusNATMAT04,kangSCI06,peurlaPRB07} that precipitates of nanometer size, most notably small particles of BaZrO$_3$ (BZO), were able to drastically improve the dc properties of YBa$_{2}$Cu$_{3}$O$_{7-\delta}$ (YBCO) in a magnetic field: critical currents and irreversibility lines were greatly raised, to the point that large scale applications could be envisaged.\\
One should mention that most studies have been devoted to the investigation of the magnetization or the dc transport properties, thus investigating the motion of flux lines over large distances or -equivalently- long time scales. A study of the very-short-distance vortex motion is a stringent probe to elucidate the strength and nature of the pinning potential, since in that regime only interactions between single vortices and single pinning centers are probed. In fact, it has been reported that even when vortices were strongly pinned with respect to the dc properties, e.g. by columnar defects, the reduction of the losses observed at high driving frequency (microwaves) did not exceed 30\%,\cite{silvaIJMPB00} and in some cases the microwave dissipation raises.\cite{wosikAPL99}\\
In a recent Letter,\cite{pompeoAPL07} we have shown that the beneficial effect of BZO inclusions on vortex pinning could be extended to the very high frequency regime (microwaves), where vortex oscillations are of the order of $\sim$1\AA,\cite{TomaschPRB88} thus indicating that the pinning wells became rather narrow. In particular, we have shown that by introducing BZO at 7\% mol. in the target the field-induced surface resistance decreased by a factor $\sim$~3 with respect to the pure film at the same temperature and field, thus indicating that BZO acts as exceptionally strong pinning centers even at microwave frequencies.\\
Associated to the issues related to vortex pinning, there is a considerable interest in the optimization of the performances of YBCO/BZO with respect to BZO concentration: on one side higher BZO concentration can have a detrimental effect on zero-field properties such as $T_c$ and the zero-field critical current density $j_{c0}$ (e.g., Ref.\onlinecite{traitoPRB06,galluzziIEEE07}), on the other side the irreversibility line and field-dependent critical current density seem to increase with BZO concentration.\cite{macmanusNATMAT04,kangSCI06,peurlaPRB07,galluzziIEEE07} Testing the microwave properties is an even more stringent check: microwaves probe a large part of the area of the film, and in films of typical thickness of $\sim$~100-200 nm the electromagnetic field fully penetrates the film, thus probing the whole volume of the sample. Percolation effects, ``best path" issues are common in dc, while at microwave frequencies the full volume is probed. Thus, we have studied the effect of increasing BZO concentration, from 2.5\% to 7\%, on the microwave thin film surface impedance in moderate magnetic fields, $\mu_0 H < 0.8 $~T. In the following, we will show that the beneficial effect of BZO particles on the vortex state microwave response increases at high BZO concentrations, and that the data do not exclude further optimization. Thus, we confirm and reinforce our previous result,\cite{pompeoAPL07} with the indication that there might be additional room for improvement of the pinning properties in YBCO/BZO.\\
The paper is organized as follows. In Section II, we  briefly outline the film growth and characterization, and we shortly describe the microwave experimental technique. The microwave data and a discussion are found in Section III. Some conclusions are drawn in Section IV.
\section{Experimental section}
\label{exp}
\begin{figure}[htb]
\begin{center}
\includegraphics [width=6.5cm]{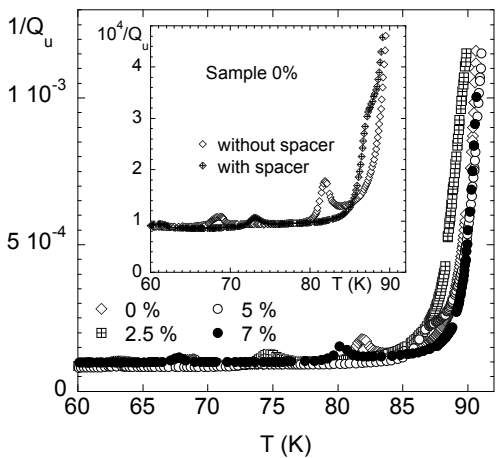}
\caption{Main panel: dependence of the absolute losses on the temperature for the samples with different BZO content in zero applied field. The figure reports the inverse quality factor $1/Q$ of the resonator with different films mounted. The BZO content does not significantly affect the transition curve in zero field. As common with dielectric resonator techniques, approaching the superconducting transition the large losses cause a strong reduction of the sensitivity, so that measurements cannot be taken further with any reliability. Peaks are due to resonances in the SrTiO$_3$ substrate, as discussed in the text and in Ref. \onlinecite{pompeoSUST07}. Inset: with the interposition of a dielectric spacer it is possible to shift the substrate resonances in order to have access to the true surface impedance.
\label{QT}}
\end{center}

\end{figure}

YBCO thin films were epitaxially grown on (001) SrTiO$_3$ (STO) substrates by pulsed laser ablation deposition technique. The technique reliably yields high-quality, epitaxial films, with narrow transition widths, high critical temperatures and high critical current densities.\cite{galluzziIEEE07,augieriJPCS08} In this work we have used pure YBCO targets for the pure, reference films, and targets with various concentration of BZO powder. Composite YBCO/BZO targets were obtained by mixing submicrometric YBCO and BZO powder and then sintering, resulting in targets with nominal content of 2.5 mol.\%, 5 mol.\% and 7 mol.\% of BZO powder. The size of the BZO grains in the target, representing an upper limit for the size of the inclusions in the films, was below 1 $\mu$m, as estimated by SEM and optical microscope analysis. X-ray $\omega$ scans through the (005) YBCO reflection revealed $c$-axis oriented films with very narrow rocking curves for all samples (FWHM $\Delta\theta < 0.2^{\circ}$, as detailed in the Table). The BZO crystallites in the YBCO/BZO films are (\textit{h}00) oriented, indicating epitaxial relationship between BZO and YBCO. SEM microscopy revealed a very smooth surface in YBCO/BZO films. Further details have been given elsewhere.\cite{galluzziIEEE07,pompeoJPCS08} Films had typical thickness of $\sim$120$\pm$10 nm. The thickness of selected samples was measured by a stylus profilometer in several locations of the films, and it was found uniform within 5\%. For all our samples, the critical temperatures as estimated from the microwave data were within the range 89.5 K - 90.6 K (see the Table). Measurements of the dc resistivity in samples grown under the same conditions confirmed that, in contrast to previous reports,\cite{traitoPRB06,galluzziIEEE07,ichinoseSUST07} but in agreement with results from other groups,\cite{hauganIEEE07} the critical temperatures did not significantly decrease with increasing BZO concentration. Critical current densities $J_c$ as measured at 77 K consistently yielded a weaker magnetic field dependence as BZO concentration was increased.\cite{augieriIEEE08} In dc, $\rho(100 K)$ changed by not more than 25\% with BZO content. In the Table we report the main parameters at various BZO content. The data in dc have been obtained in samples grown under the same conditions as those here measured. \\
Aim of this work is to investigate the pinning properties at high frequency. The presence of BZO particles induces $c$-axis elongated defects, possibly originating from self-alignment of BZO crystallites.\cite{macmanusNATMAT04,kangSCI06,galluzziIEEE07} Applying a repeated wet-chemical etching as suggested in Ref.\onlinecite{huijbregtsePRB00}, we found etch pits characteristic of linear defects parallel to the $c$-axis, with uniform distribution and areal density up to 30-40 times higher than in pristine YBCO.\cite{augieriIEEE08} Thus, apart other possible effects, our YBCO/BZO samples certainly contains more defects than pure YBCO, and those defects are elongated along the $c$-axis.\\
\begin{table}[]
\begin{tabular} {cccccc}
\hline
\%BZO&$T_{c0,min}$(K)& $T_{c0,max}$(K)  & $\rho_{100}$($\mu\Omega$cm) & $\Delta\omega$ & $T_{\mu w}$(K)\\
%
\scriptsize{(at mol.)}& & & & &\\
%
%
%
0 & 90.0 & 91.0 & 100  & 0.16$^{\circ}$& 90.4\\
%
%
2.5 & 89.2 & 90.3  & 85 & 0.12$^{\circ}$ & 89.5\\
%
%
5 & 90.0 & 90.6 & 93 & 0.09$^{\circ}$  & 90.6\\
%
%
7 & 89.7 & 90.2 & 107 & 0.08$^{\circ}$  & 90.5\\
\hline
\end{tabular}
\caption{Experimental parameters related to the characterization of the samples with different BZO content. Data taken in dc are: $T_{c0,min}$ and $T_{c0,max}$, the minimum and maximum zero-resistance temperature, respectively, as measured in several sets of samples; $\rho_{100}$, the dc resistivity at 100 K as measured\cite{augieriIEEE08} in the samples with $T_{c0,max}$. $\Delta \omega $ is the FWHM of the X-ray diffraction $\omega$ scan through the (005) YBCO reflection. $T_{\mu w}$ is the temperature where the $Q$ factor of the microwave transition reduces to 10\% of its value at 65 K (reported in Figure \ref{Q}), and is a measure of the transition temperature.
\label{tbl}}
\end{table}
Microwave measurements were performed by means of a sapphire dielectric resonator, operating at 47.7 GHz in the TE$_{011}$ mode.\cite{pompeoJSUP07} The sapphire puck was shielded by a metal plate on one side, by the superconducting film on the opposite side, and by a cylindrical metal enclosure. The latter was sufficiently far apart so not to give significant contributions to the overall response. Microwave fields and currents flowed in the $(a,b)$ planes of the samples, thus avoiding any $c$-axis contribution to the dissipation. In the present microwave field configuration, the edges of the samples played a negligible role: the microwave currents vanished at the edges of the metal enclosure, and were of maximum intensity in correspondence to the sapphire puck. The response mainly came from a circular area of $\sim$ 2 mm diameter.\cite{pompeoJSUP07}\\
Measurements of the unloaded quality factor $Q$ of the resonator are related to the surface resistance (losses): the higher the $Q$, the lower the losses. The determination of absolute surface resistance requires calibration of the resonator. In the present setup, the presence of the metal plate introduces some systematic error on the measurement of the absolute surface resistance,\cite{errors} that we estimate of the order of 10\%. By contrast, measurements of the field-induced changes in $Q$ directly report the changes in the film surface resistance $\Delta R'=R'(H,T)-R'(H,0)$. Similarly, measurements of the field- induced shift of the resonant frequency $\Delta f$ yield changes in the film surface reactance, $\Delta X'=X'(H,T)-X'(H,0)$.\\
In thin films of thickness $d\lesssim\lambda$ ($\lambda $ is the London penetration depth), such as in our case, one usually relies on the applicability of the so-called thin film approximation:\cite{silvaSUST96} $Z'=R'+\mathrm{i}X'\simeq \left(\rho_{1}+\mathrm{i} \rho_2\right)/d$, where $\rho_{1}+\mathrm{i} \rho_2$ is the complex resistivity. When field-induced variations are concerned, one has:
\begin{equation}
\begin{split}
\Delta Z'= \Delta R'+\mathrm{i} \Delta X'\simeq \frac{\Delta \rho_{1}+\mathrm{i}\Delta \rho_2}{d}
\label{thin} 
\end{split}
\end{equation}
Care must be taken when the substrate has a strong temperature dependent permittivity, as in our case with SrTiO$_3$. In this case typical substrate resonances \cite{silvaSUST96, kleinJAP90} appear at some temperature, obscuring the sample behavior. We shifted the substrate resonances to different temperatures by changing the substrate impedance with the introduction of a suitable dielectric spacer, according to the procedure described elsewhere,\cite{pompeoSUST07} thus allowing a more extensive experimental investigation. An exemplification of the effect of the SrTiO$_3$ substrate and of the insertion of a dielectric spacer is reported in the insert of Figure \ref{QT}.\\
The measurements here presented have been all taken in the thin-film regime, where Equation \eqref{thin} applies. In this case, the field induced changes in the complex resistivity of the superconducting film are obtained (without any need for a calibration of the resonator) using the standard expression:
\begin{small}
\begin{equation}
\begin{split}
\frac{\Delta \rho_{1}+\mathrm{i}\Delta \rho_2}{d}= \\
=G\left\{\left[\frac{1}{Q(H,T)} -  \frac{1}{Q(0,T)}\right]-2\mathrm{i}\frac{f(H,T)-f(0,T)}{f(0,T)}\right\}
\label{DZeff} 
\end{split}
\end{equation}
\end{small}
where $G$ is a calculated numerical factor. On the basis of Equations \eqref{thin} and \eqref{DZeff}, we note that the film thickness acts as a mere scale factor in the determination of $\Delta\rho$. In our case the film thickness is basically the same for all films, so that this scale factor does not vary more than $\pm$8\% from sample to sample.\\
Measurements were taken in all films at selected temperatures $T>$ 60 K and for fields $\mu_0H<$ 0.8 T, applied aligned with the $c$-axis (perpendicular to the film plane). We verified, by calculations and experimental check, that all measurements were taken in the linear response regime. No microwave power dependence was observed. As usually reported in microwave measurements, the field dependence is reversible, apart a small field range $\mu_0H \lesssim$~0.05 T.\\

\section{Microwave data and discussion}
\label{data}
\begin{figure}[htb]
\begin{center}
\includegraphics [width=6.5cm]{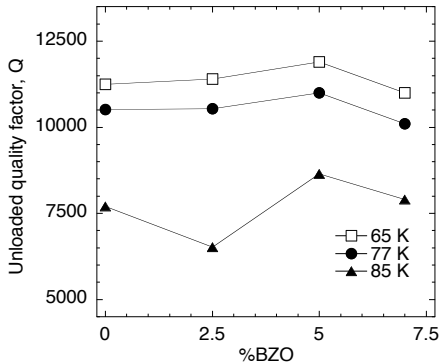}
\caption{Dependence of the absolute microwave losses on BZO content in YBCO/BZO films, in zero applied field. The figure reports the quality factor of the same resonator with different films at selected temperatures: 65 K (full circles), 77 K (squares), 85 K (diamonds). The BZO content does not significantly affect the losses in zero field at low temperatures. Only close to the transition some difference appears.
\label{Q}}
\end{center}
\end{figure}

\begin{figure}[htb]
\begin{center}
\includegraphics [width=6cm]{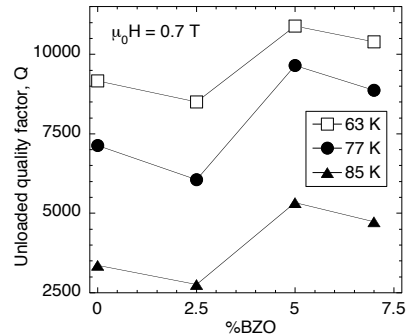}
\caption{Dependence of the absolute microwave losses on BZO content in YBCO/BZO films, with a magnetic field $\mu_0 H$= 0.7 T applied along the $c$-axis, at selected temperatures: 63 K (full circles), 77 K (squares), 85 K (diamonds). In sample with 5\% mol. BZO we could not measure $Q$ with an applied field at 63 K: the datum point reported is the average of the measures at 60 K and 65.5 K. It is evident that at higher BZO content the losses are lower (higher $Q$). Nevertheless, the BZO content dependence is different with respect to the one exhibited in Figure \ref{Q} in zero field.
\label{QH}}
\end{center}
\end{figure}

\begin{figure}[htb]
\begin{center}
\includegraphics [width=6cm]{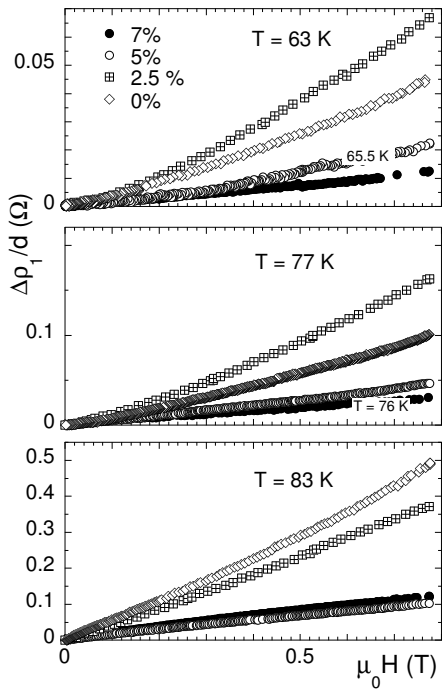}
\caption{Real part of the microwave resistivity, $\Delta\rho_1/d$, as a function of the applied field, for T = 63, 77 and 83 K (slightly different temperatures are indicated in the figure) and various BZO contents. It is seen that large amounts of BZO particles reduce significantly the field-induced losses, while small amounts (2.5\%) can increase the field-induced losses.
\label{drho1}}
\end{center}
\end{figure}

\begin{figure}[htb]
\begin{center}
\includegraphics [width=5.5cm]{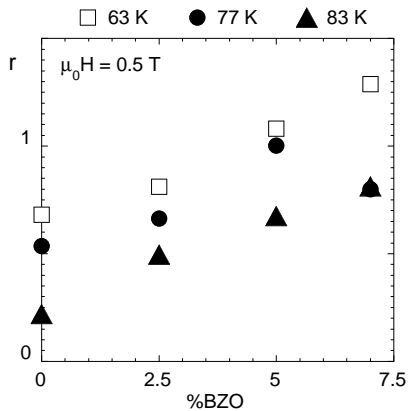}
\caption{Pinning parameter $r$ taken at $\mu_0H =$0.5 T as a function of BZO content and for $T=$ 63, 77 and 83 K (as in Figure \ref{QH}, the datum point for the 5\% BZO sample is the average of the measures at 60 K and 65.5 K). It is seen how pinning steadly increases with BZO content.
\label{rBZO}}
\end{center}
\end{figure}

\begin{figure}[htb]
\begin{center}
\includegraphics [width=6cm]{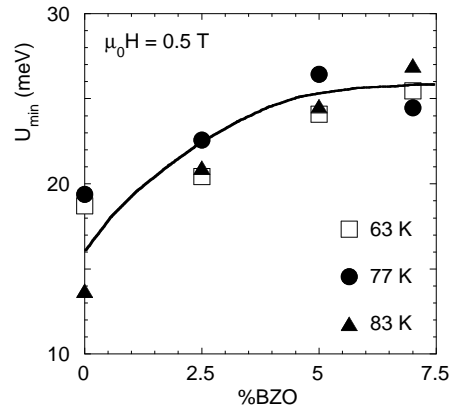}
\caption{Minimum allowed values for the barrier height of the vortex potential landscape, $U_{min}$, calculated at $\mu_0H =$0.5 T as a function of BZO content and at $T=$ 63, 77 and 83 K (as in Figure \ref{QH}, the datum point for the 5\% BZO sample is the average of the measures at 60 K and 65.5 K). It is seen that $U_{min}$ depends on BZO content and saturates at 5\% mol. BZO.
\label{Umin}}
\end{center}
\end{figure}

We first investigate the effect of the addition of BZO particles in the target on the zero-field microwave losses. We remind that in thin films the losses arise from a combination of radiation losses through the film (thickness dependent), the intrinsic (e.g., due to quasiparticles) losses, extrinsic losses due to grain boundaries,\cite{marconPRB89,wosikPRB93,halbritterJS95} metal or insulating outgrowth, etc. Absolute microwave losses are represented by the unloaded quality factor of the resonator with the sample mounted, $Q$: the higher is $Q$, the lower are the losses.\\
The superconducting transition is well characterized by the inverse quality factor $Q^{-1}(T)=A+R^{\prime}(T)/G$, where $A$ is an additional calibration factor of the resonator that may depend only weakly on the temperature, but is the same for all the measurements.\cite{pompeoJSUP07} In Figure \ref{QT} we report the transitions of the four samples investigated. As it can be seen, the BZO content has very little effect on the shape and width of the transitions, testifying the basic equivalence of the various samples in zero field. When the normal state is approached, $Q$ suddenly decreases and the resonance cannot be reliably measured throughout the transition. This is a well-known drawback of the dielectric resonator technique.\cite{pompeoJSUP07} In order to evaluate the transition temperature, we reported in the Table the temperature where $Q$ drops to 10\% of its value at 65 K.\\
We stress that the minor differences between samples have very little effect at low temperatures, where one can better find out the contribution due to BZO. \\
In Figure \ref{Q} we report measurements of $Q$ as a function of the BZO content at various significant temperatures, in zero applied field. As it can be seen, the absolute microwave losses in YBCO/BZO films depend weakly on the BZO content for $\mu_0H$=0: at 65 K and 77 K, $Q$ vs. the BZO content is nearly flat, with a possible \textit{reduction} of the losses (increase of $Q$) with moderate BZO content and a beginning of degradation at 7\% mol. The data at 85 K exhibit some more scattering, due to the finite width of the transition, but still do not show a definite worsening with increasing BZO content (we note that in the datum point at 2.5 \% and 85 K the zero-field losses might be affected by the slightly closer beginning of the transition with respect to the other samples).\\
We can conclude that the BZO content is not a significant source of deterioration of the zero-field microwave losses in the technologically interesting temperature range 65-77 K. From the dependence of the losses on the BZO content we can also infer that the microstructural properties of our films do not degrade upon moderate BZO addition, since if they did, losses would have been affected.\\
The effect of the field on the losses is first analyzed qualitatively by means of Figure \ref{QH}, where we report the data for $Q$ taken at the same BZO content as above and at similar temperatures, but upon application of a magnetic field $\mu_0H$=0.7 T. As a general trend, it is clear that the in-field absolute losses decrease with high BZO content, although a small amount of BZO content is not particularly beneficial. This is an indication of a much stronger vortex pinning, effective also at our very high driving frequency. In the following, we further study the change of strength of the vortex pinning subsequent to the introduction of BZO.\\
Using Equation \eqref{DZeff}, we extract the vortex state complex resistivity from measurements of the quality factor and of the frequency shift, following the procedure adopted previously.\cite{pompeoAPL07} Typical data for the real resistivity (losses) are reported in Figure \ref{drho1}. We note that there is a clear trend with respect to the BZO content. First, we confirm the striking result\cite{pompeoAPL07} that the addition of a noticeable percentage of BZO decreases the losses by a large factor with respect to the pure sample. Depending on the temperature, losses in the samples with BZO at 5\% mol. and at 7\% mol. are smaller by a factor larger than 3, and in some cases a reduction of 5 is observed. Moreover, a second interesting result is that the losses in the sample at 2.5 \% mol. BZO are systematically larger than in the pure sample. This behavior should be contrasted with the zero-field losses reported in Figure \ref{Q}, where a slight but appreciable reduction of the losses in zero field was observed (for temperatures not too high) with 2.5\% and 5\% BZO concentration. Thus, it seems that the introduction of BZO particles affects the microwave losses of YBCO/BZO films in two different ways: first, as can be deduced from Figure \ref{Q}, there is a reduction of the film losses in zero field upon inclusion of moderate content of BZO particles, followed by a decrease when the BZO content exceeds 5\%. Second, as suggested by Figure \ref{QH} and exemplified in detail in Figure \ref{drho1}, upon inclusion of BZO particles there is first a worsening of the field-dependent losses, and with further addition of BZO particles an extraordinarily large reduction of the losses themselves.
We propose a consistent scenario for these findings. First, the introduction of BZO particles, even at low concentration (2.5\%), increases the quality of the microstructure. As a consequence, the losses in zero field do not raise, but show instead a tendency to a reduction. Such a speculation is consistent with SEM and AFM observations, and with preliminary X-ray diffraction analysis: all those probes show a noticeable and progressive reduction of the surface roughness and an increase of the crystallinity of the films.\cite{augieriJPCS08} Second, the defects originating from BZO particles act as strong pinning centers for flux lines. In a nonzero magnetic field, when flux pinning is the key factor in the determination of the losses, the data in Figure \ref{QH} indicates that small BZO quantities (2.5\% mol.) do not give strong effects on pinning, while larger concentrations (5\% and 7\%) definitely increase flux pinning.\\
In order to further discuss the high-frequency pinning properties of YBCO/BZO films, we make use of a general model for the fluxon motion resistivity. We adopt here the well-known Coffey-Clem (CC) model \cite{cc}, whence:
\begin{equation}
    \Delta\rho_1+\mathrm{i}\Delta\rho_2=\rho_{ff}\frac{1+\epsilon\left(\frac{\nu_0}{\nu}\right)^2+ \mathrm{i} (1-\epsilon)\frac{\nu_0}{\nu}}{1+\left(\frac{\nu_0}{\nu}\right)^2}.
    \label{eqCC}
\end{equation}
where the flux flow resistivity $\rho_{ff}$ depends on the density of states and quasiparticle relaxation rate in the vortex core,\cite{BS,kopninRPP02} the creep factor $\epsilon$ is a measure of the height of the pinning potential ($\epsilon=$~0 corresponds to no creep), $\nu_0$ is a characteristic frequency and $\nu_0\rightarrow \nu_p$ for $\epsilon\rightarrow 0$, and the pinning frequency $\nu_p$ is related to the steepness of the pinning potential. We note that the analysis on the basis of different models, such as the Brandt model,\cite{brandtPRL91} does not yield qualitative differences.\cite{pompeoPRB08}\\
We focus here on the pinning-related parameters. A very useful, and often employed,\cite{halbritterJS95} experimental parameter to describe pinning is the ratio $r$, defined as:
\begin{equation}
    r(H) = \frac{\Delta\rho_2(H)}{\Delta\rho_1(H)}
    \label{eqr}
\end{equation}
This parameter is a measure of the relative weight of the reactive (elastic stored energy) and resistive (dissipated energy) response, and has the important advantage that it can be calculated directly from the raw data without any need for a calibration of the resonator [see Equation \eqref{DZeff}]. When $r \gtrsim $1 the response is mostly due to pinned vortices.\\
Interestingly, one can exploit\cite{pompeoPRB08, pompeoJSUP07b} the analytical properties of the Coffey-Clem model\cite{cc} for fluxon motion to obtain that
 $\epsilon \leq \epsilon_{max}=1+2r^2-2r\sqrt{r^2+1}$, where $\epsilon=\left[I_0(U/2k_BT)\right]^{-2}$ ($I_0$ is the modified Bessel function of order zero) and $U$ is the height of the potential barrier. Thus, from $\epsilon_{max}$ one gets a lower limit $U_{min}$ for the height of the potential barrier on a purely experimental basis. We will discuss our data in terms of $r$ and $U_{min}$.\\
\begin{figure}[htb]
\begin{center}
\includegraphics [width=6cm]{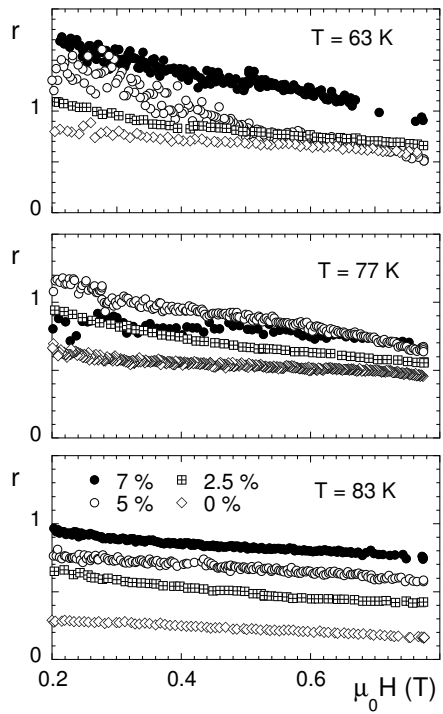}
\caption{Pinning parameter $r$ as a function of the applied field at different temperatures and various BZO contents. Data are reported above $\mu_0H=$0.2 T to avoid crowding due to numerical scattering. Slightly different temperatures are as in Figure \ref{drho1}.
\label{r}}
\end{center}
\end{figure}
The $r$ parameter is reported in Figure \ref{r}, as a function of the applied field and for various BZO content. It is cleary seen that the introduction of BZO particles leads to an increase of vortex pinning. In particular, the increase of pinning is evident at all temperatures, including temperatures close to the transition. Interestingly, pinning increases monotonously with BZO content up to 5\%, and in some cases also above 5\%. From this figure it appears that at some temperature there may be still room for improvement of the pinning properties, in particular from 83 K and below. To elucidate further the role of BZO particles, we report in Figure \ref{rBZO} the pinning parameter $r$, taken at an applied field $\mu_0H =$~0.5 T, as a function of BZO content at different temperatures. The overall trend clearly shows that pinning increases with BZO content. Thus, even if the field-induced losses are not monotonous with respect to BZO content (the sample with 2.5\% mol. BZO exhibits larger losses), the pinning-related dynamics does exhibit a monotonous behavior with respect to BZO content. We conclude that, in addition to the increased vortex pinning of interest here in view of its relevance for applications, also the flux flow resistivity $\rho_{ff}$ (i.e., the density of states and/or the quasiparticle scattering rate within the vortex cores) is most likely affected by BZO particles. This aspect will be the subject of further studies. We now present further considerations on the increased pinning properties of YBCO/BZO.\\
Additional insight can be gained from the investigation of $U_{min}$. In Figure \ref{Umin} we report $U_{min}$, the minimum height of the vortex energy landscape, calculated at $\mu_0H =$0.5 T as a function of the BZO content at three selected temperatures. It is clearly seen that the minimum height of the potential wells increases as BZO content increases and saturates at 5\% mol. BZO. It is believed that in high-$T_c$ superconductors the vortex energy landscape is very uneven, with a distribution of potential wells. By using high-frequency driving current, such as our microwave probe, oscillations of the vortices are so small that they do not escape from the potential well because of the driving current, and single-vortex response is essentially probed. In our evaluation $U_{min}$ is the lower limit for the height of the potential well, compatible with our data. Our result implies that the minimum allowed value for the energy barriers increases with introduction of BZO.
A natural explanation is that each vortex finds a preferential position on a single BZO particle. At intermediate BZO content there are still fluxons pinned by natural, weak defects. With increasing BZO content each flux line finds a pinning well corresponding to a BZO particle. This is consistent with the minimum height $U_{min}$ that saturates at 5\% mol. BZO: at $\mu_0H=$~0.5 T the areal density of flux lines is of the order $\sim \mu_0H/\Phi_0\simeq$~250 $\mu$m$^{-2}$, where $\Phi_0=$~2.07$\cdot$10$^{-15}$ Tm$^{2}$. This number should be compared to the number $n_d$ of BZO-induced defects. In our 5\% sample, using the etching method described in Section~\ref{exp}, we found\cite{augieriIEEE08} $n_d\sim$ 600 $\mu$m$^{-2}$, in good agreement with the fact that $U_{min}$ from our data does not increase above 5\% mol. BZO: all flux lines are already pinned in artificially-induced defects. This effect is reminiscent of the ``matching effect" observed in artificially patterned or in heavy-ion irradiated samples, most probably with a different efficiency in terms of the number of pinning centers: due to the random distribution of BZO-induced pinning centers, we cannot expect that exactly each defect will pin one vortex.  To further investigate this issue, measurements at higher fields or with intermediate BZO content are prospected.

\section{Conclusions}
\label{conc}
We have presented an experimental investigation of the pinning properties of YBCO/BZO films at the high edge of the microwave spectrum. A systematic analysis of films with various BZO contents, up to 7\% mol. in the target, was performed. We have found that even at our very high measuring frequency the pinning wells introduced by BZO are steep enough to pin strongly the flux lines, with an effect larger than in columnar-irradiated samples. We have found that in a moderate magnetic field the BZO content reduces the overall losses. We have analyzed our data within a mean-field model for the vortex motion. The analysis of the pinning-related parameters shows that the overall vortex pinning monotonously increases with BZO content. The minimum height of the vortex potential well increase with BZO content, and saturates at 5\% mol. Corroborated by preliminary structural investigations, we inferred that vortices are pinned by individual pinning centers, introduced by BZO particles. BZO particles are thus demonstrated to be very effective pinning centers also for high-frequency applications in a magnetic field.

\section*{\textsc{Acknowledgments}}
We are grateful to T. Petrisor and L. Ciontea at Technical University of Cluj (Romania) for target preparation.

\end{document}